\documentclass[10pt,twocolumn,a4paper,fleqn]{article}

\usepackage[numbers,sectionbib,square]{natbib} 

\usepackage{balance}
\newcommand{\myemail}{{\let\thefootnote\relax\footnote{$\star$
    \texttt{Lukasz.Bratek@pk.edu.pl}}}}
\newcommand{\mytitle}[1]{\begin{quotation}{\bf\huge\noindent #1}
    \end{quotation}}
\newcommand{\myabstract}[1]{\begin{quotation}\noindent{\bf Abstract.}{
    \small #1}\end{quotation}}
\newcommand{\mykeywords}[1]{\begin{quotation}\noindent{\bf Keywords:}{
    \small #1}\end{quotation}}    

\usepackage[utf8]{inputenc}
\usepackage[T1]{fontenc}

\usepackage{amsmath,amssymb}
\usepackage[scriptsize,bf]{caption}
\usepackage{float}
\usepackage{placeins}
\usepackage{xspace}

\newcommand{\sech}{\mathrm{sech}}
\newcommand{\ud}{\mathrm{d}}
\newcommand{\br}[1]{\left(#1\right)}
\newcommand{\ef}{\mathfrak{e}}
\newcommand{\mf}{\mathfrak{m}}
\newcommand{\micro}{\textnormal{\textmu}}
\newcommand{\muGs}{\,\micro{\rm Gs}}
\newcommand{\Gs}{\,{\rm Gs}}
\newcommand{\eV}{\,{\rm eV}}
\newcommand{\GeV}{\,{\rm GeV}}

\begin{document}
\twocolumn[\begin{@twocolumnfalse}

\bigskip\bigskip\bigskip

\mytitle{Magnetic Monopoles: Theoretical Insights into the Cosmic Ray Conundrum}

\medskip

\begin{center}
{\large  {\L}ukasz Bratek$^{*}$, Joanna Ja{\l}ocha}
\\
\medskip
\begin{tabular}{@{}l}
{\small Department of Physics, Cracow University of Technology,  
Podchor\k{a}{\.z}ych 1, PL-30084 Krak{\'o}w, Poland}
\end{tabular}\\
\bigskip \texttt{Preprint v1: proc. 7th International Symposium on Ultra High Energy Cosmic Rays (UHECR2024)
 17-21 November 2024, Malargüe, Mendoza, Argentina\\}
\end{center}

\myabstract{
Ultra-high energy (UHE) photons above $10^{18}\eV$ serve as valuable probes of fundamental physics. While typically produced in interactions involving charged particles, they could also originate from exotic sources such as annihilations of magnetically charged monopole-antimonopole pairs or decays of highly accelerated monopoles ($\sim 10^{21}\eV$). Detecting such photons would impose constraints on monopole properties. Despite strong theoretical motivations and extensive experimental searches, no monopoles have been observed to date. 
A possible explanation beyond high monopole masses arises from Staruszkiewicz’s quantum theory of infrared electromagnetic fields. His argument, rooted in the positivity of the Hilbert space norm, suggests that isolated magnetic monopoles may not be physically realizable. If correct, this would imply that while monopoles remain mathematically well-defined within field theories, only magnetically neutral configurations could exist in nature.
}

\mykeywords{magnetic monopoles, zero-frequency fields, charge quantization, quantum theory of the electric charge, infrared electromagnetic fields, 
quantum field theory on de Sitter space,
 fine structure constant, ultra high energy photons, existence of magnetic monopoles}

\renewcommand{\thefootnote}{\fnsymbol{footnote}}  
\bigskip
\end{@twocolumnfalse}]
\renewcommand{\thefootnote}{\fnsymbol{footnote}}  
\footnotetext[1]{\texttt{Lukasz.Bratek@pk.edu.pl}}
\setcounter{footnote}{0}                          
\renewcommand{\thefootnote}{\arabic{footnote}}    


Magnetic monopoles have long been hypothesized as fundamental particles that could profoundly influence ultra-high-energy (UHE) cosmic rays, potentially explaining energies beyond $10^{20}\,\mathrm{eV}$ and yielding distinctive photon signatures. Although theories---from Dirac’s pointlike charges to solitonic non-Abelian monopoles---predict them, no experimental detection exists. Recent quantum arguments also question whether free monopoles can exist. For instance, Herdegen \cite{herdegen1993} showed that excluding monopoles is crucial for consistent angular momentum in massive charged scattering, while Staruszkiewicz’s quantum theory of electric charge and phase \cite{AStar1989a,AStar1998a} suggests magnetic monopoles may violate Hilbert-space structure.

We now outline cosmic-ray phenomenology, UHE photon signatures, monopole models, and arguments against their existence; see \cite{bratek2022monopole} for further detail and references.

\section{UHECR signatures of magnetic monopoles}

Protons above $5\times10^{19}\,\mathrm{eV}$ lose energy through interactions with background radiation, limiting their range to a few megaparsecs. Heavier ions similarly attenuate, enforcing the GZK cutoff, so cosmic rays above $10^{20}\,\mathrm{eV}$ from tens of megaparsecs should be unobservable. This cutoff may also reflect source limits, since no confirmed mechanism surpasses $10^{20}\,\mathrm{eV}$. If higher-energy events are found, more exotic acceleration must be invoked.

Large-scale magnetic fields can produce the highest-energy cosmic rays. In second-order Fermi acceleration, repeated scattering off magnetized plasma can raise energies to $E_F \approx Z\beta c e B L$ before escape. Shock acceleration often reaches $10^{21}\,\mathrm{eV}$ near pulsars. A magnetic charge would enhance energy gain further.

In 1960, Porter \cite{porter1960} proposed that $\sim10^{-14}$ of all primary cosmic rays might be monopoles. Unlike Fermi processes, such monopoles could draw energy directly from magnetic fields, possibly forming in neutron-star interiors or emerging via Schwinger-like tunneling in intense fields. ’t~Hooft–Polyakov monopoles demand even stronger fields. Heavy-ion collisions at RHIC and the LHC reach $10^{18}$–$10^{20}\,\mathrm{Gs}$, exceeding $\sim10^{15}\,\mathrm{Gs}$ in magnetars. Despite searches (MACRO, Baikal, Amanda-II, RICE, SLIM, ANITA, IceCube, Pierre Auger), only upper bounds on fluxes exist.

Photon emission by magnetic monopoles is enhanced by $\sim4692$  compared to a unit electric charge, growing as $n^2$ if the charge is $n$ times the minimum,  see equation \eqref{eq:multipl}. This strong coupling boosts ionizing power, aiding detection (e.g., MoEDAL at the LHC). Astrophysical monopoles can exceed $10^{21}\,\mathrm{eV}$, above the $10^{20}\,\mathrm{eV}$ UHECR limit, largely independent of mass. A high-energy monopole traversing matter may emit a few UHE photons, detectable above $10^{19}\,\mathrm{eV}$, and monopolium states can annihilate into energetic photons.

Dirac showed \cite{dirac1948} that a monopole of mass $m$ and charge $g$ obeys
$m\ddot{x}_{\mu}{=}g G_{\mu\nu}\dot{x}^{\nu}$,
mirroring the Lorentz force law for electric charges (here, $G_{\mu\nu}{\equiv}\frac{1}{2}\epsilon_{\mu\nu}^{\phantom{\mu\nu}\alpha\beta}F_{\alpha\beta}$). With minimal losses, a magnetic field accelerates a monopole like an electric field does a charge. For Dirac’s minimal charge \(g=\tfrac{e}{2\alpha}\) over distance \(L\), the energy gain
$\Delta E= \tfrac{ec}{2\alpha}BL$
is mass-independent, exceeding typical first-order Fermi acceleration by \(\sim2\beta\alpha\):
$\Delta E{\approx} \tfrac{BL}{B_oL_o}\cdot2.054{\times}10^{21}\,\mathrm{eV}$,
$B_oL_o{=}10^{17}\,\mathrm{Gs}\cdot\mathrm{cm}$  (corresponding to $3.241{\times} 10^4\muGs{\cdot}{\rm pc}$ or $6.685\times10^3 \Gs{\cdot}{\rm AU}$).
In pulsar magnetospheres (\(B\sim10^{12}\,\mathrm{Gs},\,L\sim1\,\mathrm{km}\)), \(\Delta E\sim2\times10^{21}\,\mathrm{eV}\) matches the highest cosmic-ray energies. Extrapolating to larger scales (galaxies, clusters) yields $10^{19}$–$10^{24}\,\mathrm{eV}$ accelerations, see table \ref{tab:entable}.
\begin{table*}[h]
\centering
\renewcommand{\arraystretch}{0.7}
\setlength{\tabcolsep}{4pt}
\footnotesize
\begin{tabular}{|l|c|c|c|}
\hline
acceleration medium & magnetic field $B$ & distance scale $L$ & energy gain \\
\hline\hline
interplanetary space & $50\,\muGs$ & $1{\,\rm AU}$ & $1.5{\times}10^{13}\eV$\\
Sun-spots  & $10^3\,{\rm Gs}$ & $10^{4}{\,\rm km}$ & $2.1{\times}10^{16}\eV$\\
 Galaxy: interstellar & $2\,\muGs$ & $100\,{\rm pc}$ & $1.3{\times} 10^{19}\eV$\\
white dwarfs  & $5{\times}10^6\,{\rm Gs}$ & $10^{4}{\,\rm km}$ & $1.0{\times}1 0^{20}\eV$\\
radio-galaxy lobes & $10\,\muGs$ & $10\,{\rm kpc}$ & $6.3{\times} 10^{21}\eV$\\
clusters of galaxies  & $1\,\muGs$ & $100\,{\rm kpc}$ & $6.3{\times}10^{21}\eV$\\
active galactic nuclei  & $10^4\,{\rm Gs}$ & $5{\,\rm AU}$ & $1.5{\times} 10^{22}\eV$\\
pulsar magnetosphere  & $5{\times}10^{12}\,{\rm Gs}$ & $10{\,\rm km}$ & $1.0{\times}1 0^{23}\eV$\\
intergalactic medium  & $10^{-2}\,\muGs$ & $3{\,\rm Gpc}$ & $1.9{\times}10^{24}\eV$\\
\hline
\end{tabular}
\caption{\label{tab:entable}Selected estimates of monopole energy gains in various magnetic environments.}
\end{table*}
Radiation mechanisms for UHE monopoles mirror those of relativistic charges: a Lorentz-transformed field $\gamma cB$ causes energy losses $\tfrac1{2\alpha}\!\approx\!68$ times that of a charged particle. Monopole decay or annihilation can emit photons above $10^{21}\,\mathrm{eV}$. Despite these channels, no monopoles have been directly detected.
Their existence remains an open question, constrained by high-energy observatories. For overviews of searches, see \cite{giacomelli2005}, and for theoretical/experimental constraints, see \cite{milton2006,mavromatos2020,shnir2005,bratek2022monopole}.

\section{A magnetic monopole theory short revision}

In classical Maxwell electrodynamics, monopoles are structurally absent. Still, minor quantum topological modifications permit their introduction.
Dirac’s monopole theory \cite{dirac1931} starts with the fact that a wave function’s phase is only fixed up to a constant, allowing shifts around closed loops. He required such shifts to be universal, coupling them to the electromagnetic potential via $\frac{e}{\hbar c}\int A_{\mu}\,\ud{x}^{\mu}$. Non-integrable phase implies nonzero flux, signaling a monopole.
For a Maxwell field, any closed-surface flux vanishes. With a monopole, flux is nonzero, forcing $A_{\mu}$ to be singular. Such singularities arise from $\kappa_{\mu}=\partial_{\mu}\kappa$, where $\kappa$ is an indefinite phase. Dirac showed these fields can be reinterpreted as potentials, validating monopoles.
Because wave function phase is unchanged by $2\pi n$ shifts, charges must be quantized. For a monopole, the flux integral obeys
$
2\pi \sum n_i + \frac{e}{\hbar c}4\pi \Phi_B=0,
$
leading to Dirac’s condition 
\begin{equation} \label{eq:Dirac_formula}
g_n{=}ng, \quad g{=}\frac{\hbar c}{2e}, \quad n{=}0, \pm1, \pm2, \dots.
\end{equation}
A Dirac string cancels the monopole’s outward flux and remains unobservable due to quantization. Though $A_{\mu}$ is singular along a semi-infinite line, gauge choices merely relocate this line without altering observables. Jackiw \cite{jackiw1985} introduced monopoles without a singular $A_{\mu}$, implying pointlike monopoles.
Dirac’s formula implies monopole–photon coupling far exceeds electron–photon coupling. The force between monopoles is
\begin{equation}\label{eq:multipl}
\left( \frac{g}{e} \right)^2 = \left( \frac{n}{2\alpha} \right)^2 \sim 4692 n^2
\end{equation}
stronger than the Coulomb force between electrons, causing vastly enhanced radiation.

Field-theoretic monopoles emerge in a unifying gauge group that breaks to $U(1)$. The $SU(5)$ GUT predicts topologically nontrivial, soliton-like solutions with quantized magnetic charge and no Dirac string, spanning masses up to $10^{16}\GeV$. 
In 1974, 't~Hooft \cite{hooft1974} and Polyakov \cite{polyakov1974} discovered stable monopole solutions in an $SO(3)$ Yang-Mills-Higgs system, avoiding Dirac’s string. The approach extends to any gauge field containing $U(1)$. Their $Q{=}1$ solution is a stable soliton whose asymptotic field matches a Dirac monopole. The coupling must satisfy $\tilde{g}=\frac{e}{\alpha}$, yielding a magnetic charge $g=\frac{e}{\alpha}$, twice Dirac’s value.

\section{Staruszkiewicz’s\ argument against magnetic monopoles}

The no-monopole argument, first proposed in \cite{AStar1998a}, emerges in the infrared limit, where charge data is encoded in slowly decaying fields that lie outside the light cone. Spatial infinity thus underpins charge quantization, as in Staruszkiewicz’s quantum theory of infrared electromagnetic fields \cite{AStar1989a}.

\subsection{Electromagnetic fields at spatial infinity}

During scattering, a charge radiates a field akin to two Coulomb fields before and after scattering—its zero-frequency part is universal, set by the initial and final four-velocities, and localized outside the light cone.

All observed charged particles have nonzero mass, so currents lie inside the light cone, implying the asymptotic field must be free and homogeneous of degree ${-}2$. One isolates this zero-frequency component via the Gervais–Zwanziger limit \cite{gervais1980}, picking out $r^{-1}$ terms in $A_{\mu}(x)$ or $r^{-2}$ in $F_{\mu\nu}(x)$.
Homogeneous fields decompose into electric $\ef(x)$ and magnetic $\mf(x)$, each solving free-wave equations on a 2+1D de Sitter hyperboloid (representing the asymptotic structure of spatial infinity in 3+1D Minkowski space): 
\begin{equation}
x^{\mu} F_{\mu\alpha} = \partial_{\alpha} \ef(x), \quad
\frac{1}{2} \epsilon_{\alpha\beta}^{\phantom{\alpha\beta} \mu\nu} x^{\beta} F_{\mu\nu} = \partial_{\alpha} \mf(x)
\end{equation}
(for a Coulomb field, $\ef=e(ux)/\sqrt{(ux)^2-(xx)(uu)}$, $\mf=0$). This division is Lorentz-invariant.
Spatial infinity is modeled by a unit de Sitter hyperboloid with the line element:
$$     \ud{s}^2= g_{ij}\ud{\xi}^i\ud{\xi}^j =\ud{\tau}^2 - \cosh^2\tau \br{ \ud{\theta}^2 + \sin^2\theta \, \ud{\phi}^2}. $$
Here, $\tau$ runs from $-\infty$ to $+\infty$, with $\theta,\phi$ on $S^2$. This curved background provides a Cauchy surface for infrared fields.

Berestetskii, Lifshitz, and Pitaevskii’s criterion $\langle \vec{E}^2\rangle \gg \hbar c/(c\Delta t)^4$ \cite{BLP1982}  tests if an EM field is classical. While static fields always appear classical, the zero-frequency field of a scattered charge must be analyzed carefully. For a Coulomb field $q/r^2$, substituting $c\Delta t = 2r$ (from the opening of the light cone) gives a classical threshold at $n \gg 2.93$, as noticed by Staruszkiewicz \cite{AStar1997}.
This implies that the elementary charge $e$ is inherently quantum. If $\alpha$ were significantly different, this criterion would fail.  
This resolves why the proton field appears classical yet remains quantized.

\subsection{Staruszkiewicz’s no-monopole argument}

In quantum field theory, the structure of the Lagrangian plays a fundamental role in determining the physical properties of a theory, including whether it leads to a well-defined Hilbert space with a positive semi-definite scalar product. 
To place this in the context of infrared fields, consider a massless scalar field 
$\psi(x)$ propagating on 2+1D de Sitter spacetime. The relevant action is given by \cite{AStar1989a}:
$$
    S =\frac{\kappa}{2} \int \sqrt{-g} g^{ij} \partial_i \psi \, \partial_j \psi\,\ud^3{x}, \quad  \kappa=\frac{\hbar^2c}{4\pi e^2}.
$$ 
The classical field $\psi(x)$ satisfies the wave equation 
$\frac{1}{\sqrt{-g}} \partial_i \left( \sqrt{-g} g^{ij} \partial_j \psi \right) = 0$ which can be solved by means of separation of variables and represented as a linear combination of  mode functions $u_{lm}(x)$ with some integration constants as coefficients.
Correspondingly, the resulting real quantum field operator $\Psi(x)$ is also expanded in a series mode functions $u_{lm}(x)$:
$$\Psi(x) = \sum_{l,m} \left[ a_{lm} u_{lm}(x) + a_{lm}^\dagger u_{lm}^*(x)
   \right],$$ $$ u_{lm}(x) =
    h_{l}(\tau) Y_{lm}(\theta, \phi).$$
Here, $Y_{lm}(\theta, \phi)$ are spherical harmonics on $S^2$,
    and $h_l(\tau)$ satisfy a Klein-Gordon-like equation with a time-dependent mass term.  The exact forms of $h_l(\tau)$'s  involving hypergeometric functions are shown in \cite{AStar1989a}.
   For consistency with the classical theory, the annihilation and creation operators \( a_{lm} \) and \( a_{lm}^\dagger \) would normally satisfy the usual {canonical commutation relations} for a bosonic field:
$ [a_{lm}, a_{l'm'}^\dagger] = \delta_{ll'} \delta_{mm'}\gamma$  for $l,l'\geqslant0$ (where $\gamma$ is a possible normalization constant), with all other commutators vanishing. However,  there is an important subtlety with the $l=0$ sector and the additive term, with important consequences for the stucture of quantum states (namely, corresponding to quantum phase and charge operators), not discussed here, that must be considered separately in the context of quantum theory of electric charge \cite{AStar1989a}.

The canonical scalar product for quantum fields on curved spacetimes is given by the Klein-Gordon inner product: $\langle u, v \rangle =\frac{ \mathrm{i}}{2} \int_{\Sigma} \sqrt{-g}(u^* \partial_\mu v - v \partial_\mu u^*) g^{\mu\nu}\epsilon_{\nu\alpha\beta} \ud{x}^{\alpha}{\wedge}\ud{x}^{\beta}$ (reveling the 
characterictic for the theory of ordinary differential equations antisymmetric Wrońskian form), where $\Sigma$ is a Cauchy surface. For our case, choosing $\Sigma$ as a surface of constant $\tau$, the inner product simplifies to: 
\begin{equation}\label{eq:KGnorm}\langle u, v \rangle = \mathrm{i} \cosh^2(\tau) \int_{S^2} \sin{\theta}\ud{\theta}\ud{\phi} \, (u^* \partial_\tau v - v \partial_\tau u^*).\end{equation}
Such defined inner product is not guaranteed to be positive definite (it is at most semidefinite: e.g., for a real \( u \), we have:
$    u^* = u $,
implying: $ \langle u, u \rangle = 0$). In order to construct a proper Hilbert space, one must restrict the solution space to a subspace with a well-defined positive norm.
This is typically done by considering only positive-frequency solutions when defining the quantum field operators.

Negative norm states are problematic because they lead to unphysical probabilities in quantum mechanics. Specifically, {\it i)} the probability interpretation of the theory fails, as probabilities can become negative or arbitrarily large;  {\it ii)} the Hamiltonian is no longer bounded from below, making the vacuum state unstable; and {\it iii)}
 unitarity, a fundamental requirement of quantum mechanics, is lost.
For these reasons, quantum field theories must be formulated to avoid ghost states, ensuring that all physical states have positive semi-definite norms. 

In de Sitter space, the key difference compared to Minkowski space is that the functions 
 $u_{lm}$ lie on a dynamically expanding background, so the metric itself is time-dependent. Because no global timelike Killing vector exists in such an expanding geometry, there is no universal time translation to define positive frequency. Each coordinate choice imposes its own notion of time evolution, thus yielding different frequency splittings and vacua. De Sitter can indeed be covered by multiple coordinate patches, each introducing a distinct candidate vacuum. Consequently, a clean Lorentz-invariant separation of positive- and negative-frequency modes is not possible, allowing different definitions of the vacuum state \cite{Birrell_Davies_1982}.
 
Ensuring a well-defined Hilbert space requires careful selection of mode functions. 
Therefore, further physical prescriptions are needed to single out a particular solution.
Staruszkiewicz suggests that a solution can be considered a positive-frequency solution if it corresponds to a positive-frequency solution of Maxwell’s equations. In this case, the required normalized form of $h_l$  for positive-frequency solutions, given by 
$u_{lm}=h_lY_{lm}$, can indeed be constructed \cite{AStar1989a}  (the positive frequency $u_{lm}$ modes are normalized to $1$ with respect to the norm defined in equation \eqref{eq:KGnorm}).

\bigskip

\noindent
{\it
Staruszkiewicz's argument against the existence of magnetic monopoles} arises in a similar mathematical  context as described for scalar fields above. 
For general fields  $F_{\mu\nu}(x)$ homogeneous of degree ${-}2$, scalars $\ef(x)$ and $\mf(x)$ (homogeneous of degree zero) can be regarded as
arbitrary 
functions defined over the unit de Sitter hyperboloid.  
They are effectively functions of $\psi$, $\theta$ and $\phi$ only (independent of the fourth radial coordinate  $\chi$ enumerating de Sitter hyperboloids outside the light cone). 
The action integral ${-}\!\int\! F_{\mu\nu}F^{\mu\nu}\ud^4{x}$ expressed in terms of such fields 
 becomes a difference 
of two identical integrals \cite{AStar1989a,AStar1998a}
\footnote{
It is seen that $g^{ik}\partial_i\ef\partial_k\ef$ and $g^{ik}\partial_i\mf\partial_k\mf$ are 
both quadratic forms with signature $({+},{-},{-})$; thus, their difference is a quadratic form with signature $({+},{+},{+},{-},{-},{-})$
the same as the signature of arbitrary Maxwell Field $F_{01}^2{+}F_{02}^2{+}F_{03}^2{-}F_{23}^2{-}F_{31}^2{-}F_{12}^2$
 (in a given 
inertial frame, the latter form can be written as a difference $\vec{E}.\vec{E}{-}\vec{H}.\vec{H}$ with a contribution from 
electric field $\vec{E}$ and from magnetic field $\vec{H}$).}
$2\!\int\! \frac{\ud{\chi}}{\chi}\frac{\sin(\theta)}{\sech^2(\psi)}\ud{\psi}\ud{\theta}\ud{\phi}
\br{g^{ik}\partial_i\ef\partial_k\ef{-}g^{ik}\partial_i\mf\partial_k\mf}$.   
Disregarding the factor 
$\ud{\chi}{/}\chi$, 
both Lagrangian densities are identical to one for a free massless scalar field on the $2{+}1$D de Sitter spacetime.  
  The action integral for this system regarded as confined  to the de Sitter unit hyperboloid $\mathcal{H}_1$ can be defined to be
\begin{equation}\label{eq:properaction}
S[\ef,\mf]=C\int_{\mathcal{H}_1}
g^{ik}\br{\partial_i\ef\partial_k\ef{-}\partial_i\mf\partial_k\mf}\sqrt{{-}g}\,\ud^3{\xi}.
\end{equation}
Here, $g_{ik}$ is the metric tensor in arbitrary intrinsic coordinates $\xi^0,\xi^1,\xi^2$ on $\mathcal{H}_1$;  $C$ is a positive dimensional constant introducing the correct absolute physical scale of the action integral.
The function $\ef$ is called the electric part, while the function $\mf$ is called the magnetic part of the field.
 Both parts are dynamically 
independent and can be investigated separately \cite{AStar1989a}.   
They evolve as massless scalars on de Sitter spacetime, satisfying the free wave equation. 

The scalars $\ef$ and $\mf$ appear to be completely independent fields, Lorentz--invariantly separated from each other.
 These fields satisfy the same wave equations as the scalar field $\psi$ discussed above and are defined by the standard Lagrangian densities for the two real scalar fields $e$ and $m$: 
$    \mathcal{L}_e = \sqrt{-g} g^{ij} \partial_{i} e \partial_{j} e$ and $
     \mathcal{L}_m = \sqrt{-g} g^{ij} \partial_{i} m \partial_{j} m$.
As we have seen, upon quantization, each Lagrangian leads to a well-defined Hilbert space with a positive semi-definite inner product, ensuring physical consistency. Now, if we consider the sum of the two Lagrangians, $    \mathcal{L}_+ = \mathcal{L}_{e} + \mathcal{L}_{m}$,
 the resulting theory describes two independent quantum fields, each contributing a positive semi-definite inner product. This corresponds to a direct sum of two Hilbert spaces, each with standard probabilistic interpretation, and with composite scalar product $\langle e_1, e_2\rangle+\langle m_1, m_2\rangle$.
However, the situation drastically changes if we consider the difference:
$\mathcal{L}_- = \mathcal{L}_{e} - \mathcal{L}_{m}$.
 This negative sign tied to \(m(\xi)\)  yields an inner product of the form
$\langle e_1, e_2\rangle - \langle m_1, m_2\rangle$. 
Here, the field \(e\) still contributes to a positive semidefinite norm, but the field \(m\) now enters with the wrong sign in its kinetic term. This indicates the presence of a ghost field, which inevitably produces negative-norm states, violating unitarity and its probabilistic interpretation. To avoid these ghost fields, quantum field theories must ensure that all physical states possess positive semidefinite norms. Negative norm states are problematic because they allow the appearance of negative or arbitrarily large probabilities, remove the lower bound on the Hamiltonian (rendering the vacuum unstable), and destroy unitarity---a cornerstone of any consistent quantum theory.

From the action integral \eqref{eq:properaction}, upon quantization, the electric component \(\ef\) yields a positive-definite norm, while the magnetic component \(\mf\) generates negative-norm states, rendering the quantum theory inconsistent. To maintain a valid Hilbert space, one must discard \(\mf\), implying magnetic monopoles, which generate such negative-norm contributions, cannot exist as isolated entities. The remedy is to set \(\mf = 0\), thereby removing the negative-norm states and ensuring a positive-definite scalar product. This procedure preserves Lorentz invariance and unitarity without necessitating arbitrary state selections.\footnote{Attempts to keep both scalar fields while discarding negative-norm states fail because Lorentz invariance requires a consistent treatment of all modes; observer-dependent state removals are not physically meaningful, and the evolution of the full quantum theory mixes different modes, so any \emph{ad hoc} choice of positive-norm states is not preserved.} Retaining only \(\ef\), although it may require careful tuning of modes, need not produce negative-norm states. However, including \(\mf\) invariably introduces negative norms, violating fundamental quantum-mechanical principles in de Sitter space. Because the kinetic term for \(\mf\) is negative, it must be excluded (\(\mf = 0\)), ruling out magnetic monopoles—or any freely propagating magnetic component—in the asymptotic Maxwell field. Even with a CP-violating \(\Theta\) term, the negative-sign piece persists \cite{AStar1998b}, so \(\mf' = 0\) and free monopoles disappear.

\section{Conclusion}

While high-energy photons from monopole-antimonopole annihilation or accelerated monopoles could constrain monopole properties, accelerator and cosmic-ray searches have revealed no evidence. Beyond large predicted masses, deeper quantum arguments suggest a fundamental conflict with quantum electrodynamics. In Staruszkiewicz’s infrared theory \cite{AStar1989a}, maintaining a well-defined action at spatial infinity eliminates the magnetic part of the zero-frequency field, forbidding free magnetic charges. Related work indicates that including magnetic fields in the de Sitter infrared regime leads to indefinite norms \cite{herdegen1993,AStar1998a}, making free monopoles incompatible with a positive-definite Hilbert space. 

\bibliographystyle{JHEP}
\bibliography{2025LBJJ_monopoles_pos}

\end{document}